\documentclass[acmsmall]{acmart}

\usepackage{subcaption}
\usepackage{array}

\AtBeginDocument{%
  \providecommand\BibTeX{{%
    \normalfont B\kern-0.5em{\scshape i\kern-0.25em b}\kern-0.8em\TeX}}}

\setcopyright{acmlicensed}
\acmJournal{PACMHCI}
\acmYear{2021} \acmVolume{5} \acmNumber{CSCW1}
\acmArticle{145} \acmMonth{April} \acmPrice{15.00}
\acmDOI{10.1145/1122445.1122456}

\copyrightyear{2021}

\begin{document}

\title{Design and Appropriation of Computer-supported Self-scheduling Practices in Healthcare Shift Work}

\author{Alarith Uhde}
\email{alarith.uhde@uni-siegen.de}
\orcid{0000-0003-3877-5453}
\author{Matthias Laschke}
\email{matthias.laschke@uni-siegen.de}
\orcid{0000-0003-2714-4293}
\author{Marc Hassenzahl}
\email{marc.hassenzahl@uni-siegen.de}
\orcid{0000-0001-9798-1762}
\affiliation{%
  \institution{Siegen University}
  \department{Business Information Systems}
  \streetaddress{Kohlbettstraße 15}
  \postcode{57072}
  \city{Siegen}
  \country{Germany}
}


\begin{abstract}
  Shift scheduling impacts healthcare workers' well-being because it sets the frame for their social life and recreational activities. Since it is complex and time-consuming, it has become a target for automation. However, existing systems mostly focus on improving efficiency. The workers' needs and their active participation do not play a pronounced role. Contrasting this trend, we designed a social practice-based, worker-centered, and well-being-oriented self-scheduling system which gives healthcare workers more control during shift planning. In a following nine month appropriation study, we found that workers who were cautious about their social standing in the group or who had a more spontaneous personal lifestyle used our system less often than others. Moreover, we revealed several conflict prevention practices and suggest to shift the focus away from a competitive shift distribution paradigm towards supporting these pro-social practices. We conclude with guidelines to support individual planning practices, self-leadership, and for dealing with conflicts.
\end{abstract}



\setcopyright{acmlicensed}
\acmJournal{PACMHCI}
\acmYear{2021}
\acmVolume{5}
\acmNumber{CSCW1}
\acmArticle{145}
\acmMonth{4}
\acmPrice{15.00}
\acmDOI{10.1145/3449219}

\maketitle

\section{Introduction}

Shift scheduling is the central process of work organization in many healthcare communities. Its main purpose is to guarantee a certain quality of care by ensuring that enough healthcare workers are present at all times. The process of scheduling itself and the resulting schedules have major implications for the involved stakeholders, i.e., the management, healthcare workers, and receivers of the care. Economic, legal, as well as social interests have to be carefully weighed up against each other to find acceptable schedules. Because of this complexity, scheduling is time consuming and demanding for the planners. Consequently, algorithmic systems are designed to assist in scheduling.

Scheduling systems typically follow an abstract logic of optimization~\cite{Noack2008}. A schedule has to meet several operational constraints, including legal regulations (e.g., breaks between shifts, required number of staff) and economic goals (e.g., cost-efficiency, risk management)~\cite{VandenBergh2013}. The scheduling process consists of matching workers with shifts while satisfying these constraints. Hence, existing shift scheduling systems optimize for efficiency both in the process and the result, aiming for minimal user interaction and ``optimal'' staff assignment (e.g.,~\cite{Constantino2011, Lin2015, Solos2013, VandenBergh2013}).

From the healthcare workers' perspective, shift scheduling impacts not only their work but also their private lives. For example, irregular or long, uninterrupted shift patterns have negative effects on health~\cite{Akerstedt1990, Brown2016, Bohle1989, Knutsson1989, Rotenstein2018, Rutenfranz1982}. Specific shift assignments, e.g., on weekends or religious holidays, disturb social life~\cite{Barton1995, Colligan1990, Costa2016, Loudoun2008, Walker1985}. Successful work within a shift often depends on the particular composition of the shift team. Thus, individual considerations (e.g., fatigue, private plans, relationships with co-workers) are important. Consequently, workers value individual control which allows them to incorporate personal considerations with the schedule~\cite{Fenwick2001, Kubo2013, Nelson2010}.

While giving workers control of the schedule has positive effects on their health and job satisfaction (e.g.,~\cite{Fenwick2001, Kubo2013, Nelson2010, Nijp2012}), it is rarely addressed explicitly by shift scheduling systems. Some systems have been extended, for example, with fairness mechanisms~\cite{Constantino2013, Constantino2011, Constantino2015, Lin2015}, but they reduce the role of workers to submitting preferences for free shifts that are then further processed by the algorithms. Mostly workers are kept out of the actual decision-making process about which preferences are granted to whom and why. On the other hand, a few studies have explored the effects of more fundamental worker control through autonomous scheduling (e.g.,~\cite{Bailyn2007, Miller1984, Roennberg2010, Wortley2003}), but they lack detailed insights based on actual, subjective experiences of workers using these systems. Informed recommendations about how individual and group-level scheduling practices in a computer-supported scheduling system should be designed for are still missing.

In this article, we present a detailed case of computer-supported worker-oriented shift scheduling. Our first goal was to design an interactive scheduling system to support decentralized, individual and group-level shift scheduling practices. Our second goal was to study how workers use and experience such a system in practice to derive insights for improving the design of worker-oriented scheduling systems. In the remainder of the article, we first present related work that grounded our study in the literature. Then we outline our design process in detail, as well as the following nine months appropriation study in a retirement community. Based on our findings of the workers' subjective experiences, we present practical design guidelines for future worker-oriented scheduling systems.

\section{Related Work}

\subsection{Shift Scheduling and Well-being}

Shift work negatively affects workers' well-being, specifically by disturbing their sleep~\cite{Akerstedt1990, Brown2016}, physiological and psychological health~\cite{Bohle1989, Knutsson1989, Rutenfranz1982}, and social life~\cite{Barton1995, Colligan1990, Costa2016, Loudoun2008, Walker1985}. Moreover, these effects may reinforce each other: Workers who are dissatisfied with their social life have a higher risk for health issues such as burnout~\cite{Hulsegge2020}, a prevailing problem among healthcare workers~\cite{Hulsegge2020, Rotenstein2018}. This negative impact can be partially mitigated by granting workers some control of their own schedule~\cite{Fenwick2001, Kubo2013}. ``Self-scheduling'' is such a more inclusive, collective planning process where workers themselves, rather than a central planner, create the entire schedule~\cite{Miller1984, Roennberg2010}. It was found to increase workers' ability to integrate their private needs with their work schedule and to lower turnover rates, while improving cooperative behavior and a positive team spirit~\cite{Bailyn2007, Miller1984, Wortley2003}.

Despite the potential benefits in terms of well-being~\cite{Garde2012, Laschke2020b, Moen2016}, self-scheduling has its own specific challenges. If the team is not aware of or ignores legal and economic regulations during planning, shifts can be over- or understaffed. Moreover, planning becomes more difficult with larger teams and lower team coherence, which can in turn interfere with successful and fair planning~\cite{Bailyn2007, Wortley2003}. Finally, some workers have favored externally imposed, fixed schedules, arguing that they make their free times more predictable~\cite{Ingre2012}. In addition, we know from other work-related decision making about, for example, room temperature control~\cite{Clear2017} and task assignments for hospital porters~\cite{Bossen2016} that inclusive processes bring their own potential for conflicts and can increase accountability and pressure on the workers. In sum, increasing workers' control of their schedule through self-scheduling can support well-being, but may require support in dealing with complex regulations, addressing difficult social situations, and to keep the plan reliable.

\subsection{Computer-Supported Shift Scheduling}

In the Human-Computer Interaction (HCI) and Computer-Supported Cooperative Work (CSCW) literature, shift scheduling has not been studied extensively. Research on hospital work organization has mostly focused on task scheduling within a shift~\cite{Bardram2010, Bossen2016, Krabbel1997, Stisen2016}, including equipment and patient transfer between different groups or institutions~\cite{Bossen2016, Kuchera2011}, time critical task management during surgeries~\cite{Bardram2010, Egger1992, Kusunoki2015}, and documentation of patient records~\cite{Moller2013, Wolf1997}. Strauss et al.~\cite{Strauss1985} have presented an early and detailed analysis of the complex interactions between different actors in hospitals across an ``illness trajectory''. They described how characteristics of the illness, staff, patients and their relatives, economic pressure, and technological affordances lead to an interwoven system and bring about partially unpredictable consequences for the actors. Fitzpatrick et al.~\cite{Fitzpatrick2013} provide a more recent and more technology-centered overview. A few articles focused on information transfer during shift handovers~\cite{Fitzpatrick2004, Tang2007, Zhou2010}, and one on self-care support for nurses to cope with individual challenges of shift work~\cite{Nunes2018}.

Concerning cooperative shift planning, Uhde et al.~\cite{Uhde2020a} compared different fairness concepts during shift-related conflict resolution. Healthcare workers valued ``equality'' for critical scheduling parameters in general, for example that everyone has the same number of free weekends. However, in case of particular conflicts, they preferred a resolution process based on personal needs over equality- or performance-based (equity) approaches (see also~\cite{Deutsch1975}). Moreover, workers found collaborative decision-making fairer than decisions made by a computer, independent of the outcome (i.e., whether or not they ``won'' the free shift). Taking a more comprehensive approach, Rönnberg and Larsson~\cite{Roennberg2010} have presented a prototypical ``self-scheduling'' algorithm. Their system allowed healthcare workers to submit a preferred schedule to their group leader and tag certain work or leisure times with different degrees of personal importance. These schedules were then integrated by the leader on a computer. In case of a conflict, i.e., when two or more workers had submitted ``vetoes'' which could not be granted simultaneously, the leader resolved the problem. Feedback on the system was mixed. While some workers saw a positive potential in computer-supported scheduling, others were afraid to lose influence on their work times, which is contrary to the central intention of self-scheduling. Notably, in this case, the system had not been created as a tool to be directly used by the workers, but was based on the group leader as a central middleman and thus somewhat similar to more centralized, automated approaches~\cite{Constantino2013, Constantino2011, Constantino2015, VandenBergh2013, Lin2015}. It was generally seen as promising in terms of efficiency, when compared with manual-only self-scheduling. However, for future development, a stronger direct involvement of workers during the design and implementation as well as a worker-centered interface design may be required~\cite{Roennberg2010, Russell2012}.

Taken together, the negative impact of shift work in general~\cite{Brown2016, Costa2016, Hulsegge2020, Rotenstein2018} and the mitigating effect of worker control more specifically~\cite{Fenwick2001, Kubo2013} have been studied extensively. Self-scheduling is a concept of high worker control~\cite{Bailyn2007, Miller1984, Wortley2003} and therefore promising to reduce the negative effects of shift work on healthcare workers. However, existing studies of technical support for self-scheduling remain abstract (e.g.,~\cite{Roennberg2010, Uhde2020a}), indicating the need for more contextualized, in-depth research. Thus, we conducted a worker-centered design process of a self-scheduling system and present it in detail below.

\section{Designing a Worker-oriented Self-scheduling System}

\subsection{Setting}

The present case was set in a medium-size retirement community in Germany, with around 45 healthcare workers and 120 residents. In this retirement community, schedules are separately created by the leaders of five different subcommunities on a monthly basis. They collect preferences for days off from their team members (including themselves) in a paper notebook. Besides their planning responsibility, the subcommunity leaders also work in care, together with their team members. Once a schedule draft is finished, it is handed to a central planner who makes a legal and economic check, possibly adapts the schedules, and returns them to the subcommunities. Unlike the subcommunity leaders, the central planner is part of the management and does not work in care. He is located in a separate office area in the same building. In line with previous findings~\cite{Roennberg2010}, two central issues and expectations were reported in a preliminary meeting with the different stakeholders. On the one hand, healthcare workers hoped for the system to give them more control of their shift assignments. On the other hand, the subcommunity leaders as well as the central planner hoped that the new system would save them time.

\subsection{Design Process}

We began the design process in May 2017 with the central goal to develop an interactive, cooperative self-scheduling system for healthcare workers and to study its appropriation in context. To that end, we framed self-scheduling as an important means to support healthcare workers' subjective well-being~\cite{Diener1985, Martela2019}. The system's goal was to establish individual and collaborative practices of shift-planning, which give healthcare workers more autonomy and control while assuring the necessary quality of the resulting schedule. Because of the clear focus on workers' well-being, we used ``Design for Well-being'' (DfW)~\cite{Klapperich2019a} as the general conceptual design approach.

DfW assumes that well-being is the consequence of engaging in meaningful or enjoyable, positive everyday (work) practices. A practice is experienced as positive if it fulfills psychological needs, such as autonomy, relatedness, or competence~\cite{Hassenzahl2010a, Klapperich2019a, Sheldon2001}. Through the way practices are structured~\cite{Reckwitz2002, Shove2012}, they can lead to more or less need fulfillment, which leads to more or less satisfactory experiences~\cite{Klapperich2019b, Klapperich2020, Laschke2020a}. Importantly, DfW assumes technology to be instrumental in shaping most practices. Therefore, technology can facilitate or obstruct well-being. Rather than focusing on problem-solving alone (e.g., ``how can a legal shift schedule be created?''), DfW encourages designers to identify and emphasize the positive potential inherent in current or future work practices (e.g., ``which parts of scheduling are especially enjoyable/meaningful and why?''~\cite{Lenz2019}). Accordingly, the central design units are positive practices which include technology (i.e., material), user skills (i.e., skill), and need fulfillment (i.e., meaning)~\cite{Klapperich2019a}. In shift scheduling, for example, a scheduling app (material) may afford the user to actively integrate both their private and work life in the schedule (skill), which can charge the activity with an experience of autonomy (need fulfillment/meaning).

To that end, our first step was to identify existing positive shift scheduling practices and their associated psychological needs. Thus, we started with interviews about specific scheduling activities that healthcare workers and planners experienced as positive in the centralized setting. In a second step, we used these positive practices as the basis for the design of new technologically mediated self-scheduling practices. Of course this involved a creative transfer process, because future self-scheduling practices are by definition different. Nevertheless, they may borrow from already existing practices. The third step was the implementation of our self-scheduling practices as an interactive tablet app including usability testing (not reported here extensively). Finally, we introduced the app in one subcommunity to study its appropriation in everyday work.

\subsubsection{Step 1: Identifying Positive Practices in Current Centralized Planning}

In order to identify current positive formal and informal practices of shift planning, we conducted five initial interviews using the Positive Practice Canvas (PPC;~\cite{Klapperich2018}). The PPC is a tool specifically designed to gather positive practices from domain experts. First, we asked for positive activities related to shift scheduling and sort them from ``most enjoyable'' to ``least enjoyable''. Second, starting with the most enjoyable practice, the interviewees further decomposed the practices into the material and skills required, as well as the psychological needs that are fulfilled through the interplay of material and skills. We employed a laddering technique~\cite{Reynolds1988}, to relate elements of a scheduling practice to underlying needs (e.g., ``creating (skill) an early overview of planned vacations (material) gives security (need)'').

The participants in our PPC interviews were staff members from all three roles involved in shift planning (i.e., workers, subcommunity leaders, central planner). Because there was only one central planner who had started his job a few months prior to the interviews, we interviewed another employee who had worked as central planner for several years in the same retirement community before but had switched to a different position in the meantime. In addition, we included two subcommunity leaders who, as reported by the general manager, particularly enjoyed planning and were thus suitable interviewees. One of them was the leader of the subcommunity where the later appropriation study was set. Given that both worked as certified nurses themselves, we included questions about their role as planner and as nurse, so that they doubled in their roles, assuring the appropriate representation of the healthcare workers' perspective in our data collection. This procedure allowed us to overcome the problem of interviewing workers with scheduling experience in a setting that did not designate any planning activities besides submitting preferences for free time to regular workers. In addition, we interviewed one healthcare worker with no planning responsibilities besides submitting his preferences, although the interview turned out to be difficult because of the healthcare worker's limited experience with shift planning. All interviews took approximately one hour and were led by the same interviewer (the first author). Two participants denied video or voice recording, but we took extensive notes on the PPC itself. The other three interviews were recorded and transcribed in German for further analysis.

The PPC interviews resulted in a total of fifteen positive practices of centralized planning. In the next step, we matched practices that had a similar function in the centralized planning system, resulting in ten distinct centralized scheduling practices (see Table~\ref{tab:10practices}). For example, both central planners had a practice of integrating the printed schedule of the previous month, which included several short-term changes, with a digital system to see who had worked how many hours. We included all needs in the integrated practice that were fulfilled for at least one participant. In this case, for example, only one planner described an experience of competence when integrating the schedule, so we included the need. The central need fulfillment for this practice were the security experienced when each worker's hours were correctly counted and the feeling to be able to help them (popularity) if otherwise forgotten working hours could be included.

In order to verify our correct understanding of both the practices and their underlying needs, we then wrote a short story of about 100 words representing each of the ten centralized scheduling practices, including their specific links to psychological need fulfillment. We described the practice in an ideal way, that is, if everything worked out perfectly. Each of these stories was then reviewed in an informal interview with one of the previous interviewees who had mentioned that practice during the PPC interviews. While the interviewees generally considered the stories as accurate, we corrected small misunderstandings based on their feedback. An example idealized practice (``Making changes on short notice'' in Table~\ref{tab:10practices}) addressing the need for popularity was:

\begin{quote}
\textit{``One of the colleagues called in sick in the evening for the next day (the afternoon shift). Apparently she feels really bad, so the ward leader assures her that he takes care of finding a replacement. He knows that two co-workers would agree to stand in, so he first tries to reach one of them to ask if she could do it. The co-worker directly signals empathy for the sick colleague and asserts that she can jump in. Thankful that the co-worker is so cooperative, he marks the change in the schedule. Finally, he sends a text message to the colleague - everything is fine, she can take her rest, the co-worker takes over.''}
\end{quote}

Notably, this idealized description strongly connects to the emotional experience from the planner's point of view. We used this to validate the associated need fulfillment with the interviewees during the reviews. After the validation of all stories, four major and one minor needs emerged as sources of positive emotions during shift scheduling:

\begin{itemize}
  \item Autonomy was experienced when a healthcare worker wrote a request for a free day in the notebook as an act of prioritizing private life
  \item Popularity/helping others was prominent in pro-social practices. For instance, the subcommunity leaders had a practice of making small changes in an otherwise finished schedule, such as swapping the morning and afternoon shift between two workers based on their knowledge of each individual's preferences.
  \item Security was the primary need concerning predictability of the schedule. For example, one of the central planners mentioned that he created an overview of vacation times of each employee early on to be able to anticipate staff shortages.
  \item Competence was experienced by both subcommunity leaders during the combinatorial challenge of fitting the initial schedule together.
  \item Finally, the minor need was stimulation, which was experienced by the central planner who was new on the job and explored the new tools. However, we considered this stimulation not to be derived directly from scheduling itself and excluded it from the further process.
\end{itemize}

\begin{table*}
\begin{center}
\scriptsize
\hyphenpenalty10000
\renewcommand{\arraystretch}{1.3}
\caption{The ten centralized scheduling practices identified in the PPC interviews. The roles are CP = ``Central Planner'', HW = ``Healthcare Worker'', and SL = ``Subcommunity Leader''. The needs are Aut = ``Autonomy'', Com = ``Competence'', Pop = ``Popularity'', Sec = ``Security'' and Sti = ``Stimulation''.}
\begin{tabular}{>{\raggedright}p{3cm}cll>{\raggedright\arraybackslash}p{5cm}}
\toprule
Practice                                     & Role & Primary Needs & Secondary Needs & Description \\
\midrule
Strategically relieving employees            & CP   & Pop           & Com, Sec, Sti   & When planning for special events (e.g., a company barbecue), making sure that workers who had shown extraordinary commitment can participate \\
Creating an overview of the year             & CP   & Sec           & Pop             & Entering all vacation requests and larger events of the next year in the calendar to anticipate staff shortages \\
Distributing the new (paper) shift schedules & CP   & Pop           & Sec, Com        & Walking to all subcommunities to distribute the schedules for the next month \\
Closing revision of the previous month       & CP   &  Pop, Sec     & Com, Sti        & Including swapped shifts and other changes to assert that everyone's working hours are correctly counted \\
Collectively create the night shift schedule & SL   & Sec           & Com, Pop        & Get together with all SLs to plan the night shifts for the following month \\
Granting a wish                              & SL   & Pop           & Sec             & Including a submitted wish in the schedule \\
Improving the shift schedule                 & SL   & Com, Sec      & Pop             & Smoothing out the schedule to make it more comfortable for the workers (e.g., swapping an early and a late shift based on one‘s knowledge of their preferences) \\
Making changes on short notice               & SL   & Pop           &                 & E.g., finding a replacement if someone calls in sick \\
Correcting the working hours                 & SL   & Com, Sec      &  Aut, Pop       & Including the changed working hours of the workers (e.g., due to swapped shifts) in the schedule so the central planner sees them \\
Submitting a wish for a free shift           & HW   & Aut           & Com, Pop        & Writing a wish for the next month in the notebook \\
\bottomrule
\end{tabular}
\label{tab:10practices}
\end{center}
\end{table*}

\subsubsection{Step 2: Designing a Self-scheduling Concept Based on the Identified Practices}

\begin{figure*}
  \centering
  \includegraphics[width=0.7\textwidth]{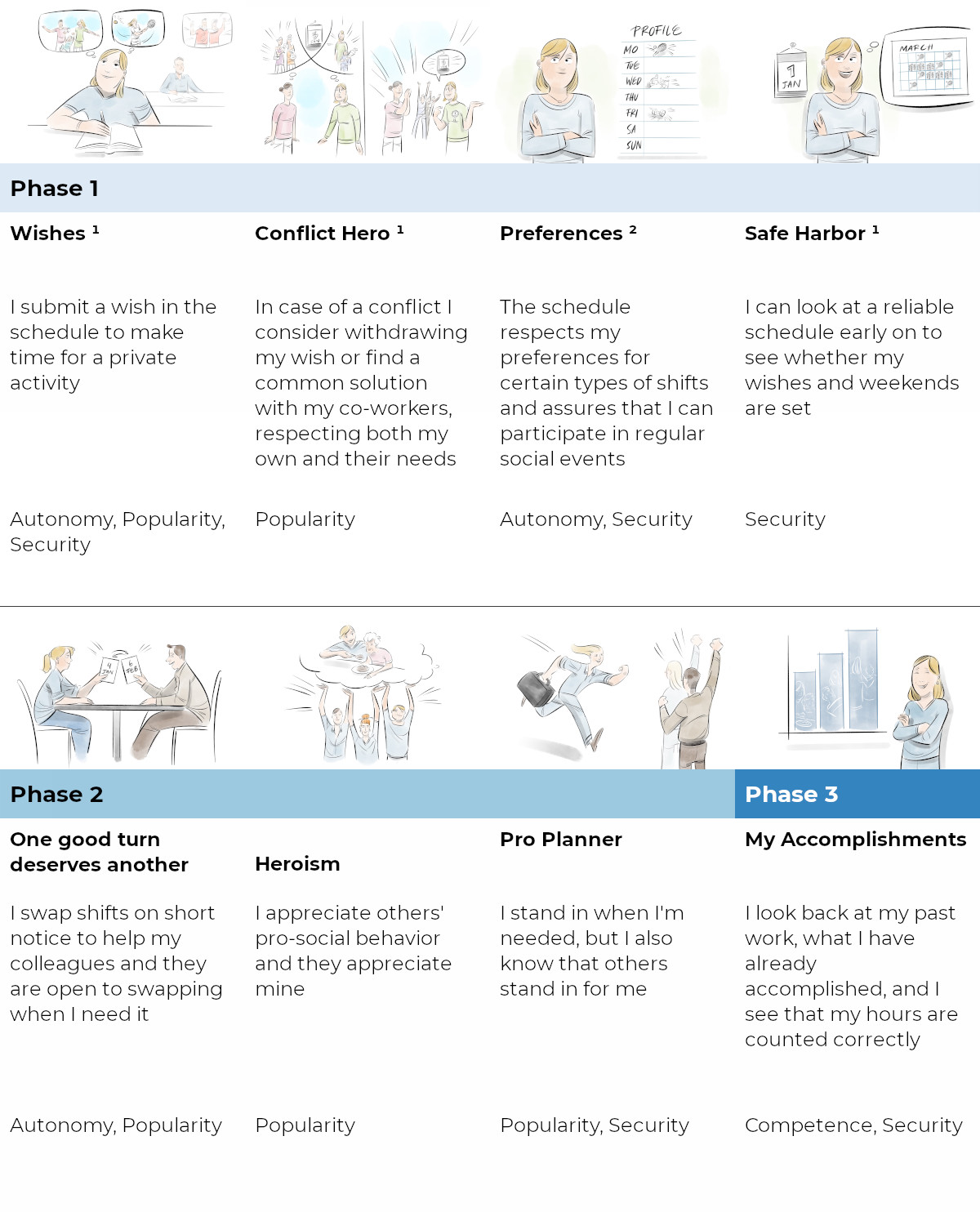}
  \Description{The figure is a user journey with the 8 positive practices of a full scheduling cycle. The cycle consists of the three phases described in section 3.2.2. Each practice is described with a positive subjective description from the perspective of a healthcare worker and the needs it addresses. Phase 1 starts with the wishes practice, which allows healthcare workers to submit a wish in the schedule to make time for a private activity and addresses autonomy, popularity, and security. Conflict hero is a practice of withdrawing one's wish to help a co-worker and addresses popularity. The preferences practice allows workers to submit general preferences, e.g., for early shifts that are automatically considered whenever possible, and it addresses autonomy and security. The safe harbor practice refers to the early availability of the schedule with accepted wishes and addresses security. In phase 2, one good turn deserves another is a practice to swap shifts on short notice, after the schedule has been published, and addresses autonomy and popularity. Heroism is a general appreciation practice for the pro-social behavior of co-workers and addresses popularity. Pro planner is a practice of standing in for the team if a co-worker is sick and addresses popularity and security. Finally, phase  3 contains the practice my accomplishments which allows workers to look back at previous shifts and check that their working hours are counted correctly. It addresses competence and security.}
  \caption{The eight positive practices forming our self-scheduling concept including the psychological needs they address. Practices marked with a (¹) were directly integrated in the app, while the ``Preferences'' practice (²) was indirectly and manually integrated by the subcommunity leaders.}\label{fig:practices}
\end{figure*}

The ten centralized scheduling practices and four needs extracted from centralized planning served as the foundation of our conceptual design of the self-scheduling process (see Figure~\ref{fig:practices}). Of course, given the fundamental change in work organization, not all of them could be used directly. For example, the practices ``granting a wish'' and ``distributing the new (paper) shift schedules'' only make sense in a centralized system. Thus, in some cases we had to replace the functions of the subcommunity leader in the current system with a collaborative process among healthcare workers. Moreover, competence played less of a role, because it was tied to the combinatorial challenge of a human planner to assemble a schedule. Given that we focused on self-scheduling practices and gave the primary influence to the healthcare workers directly, we prioritized their autonomy above a planner's need for competence. Therefore we used a computer-based mechanism to identify scheduling conflicts and to mediate the interaction on a group level. The combinatorial challenge was thus delegated to an algorithm.

The resulting scheduling process was based on eight self-planning practices for healthcare workers divided into three phases. Phase 1 starts before the schedule sets in. It contains four practices of schedule preparation. Phase 2 starts after the initial schedule is completed and accounts for changes on short notice while the schedule is already running. Phase 3 is set after the schedule was worked through and contains a retrospective practice. The next planning cycle starts prior to the next month, overlapping with the previous cycle. We presented the new process with a video prototype in two workshops, one with the subcommunity leaders and one with non-planning healthcare workers. In addition, we ran three individual interviews with non-planning healthcare workers. Feedback was generally positive and the new process was perceived as realistic and resonating with the workers' experiences. One critical remark from all three individual interviews was that the appreciation for the co-workers who stand in (see the ``pro planner'' practice in Figure~\ref{fig:practices}) was somewhat exaggerated in the video, where it was emphasized with an audible applause. Nevertheless, the appreciation itself was seen as desirable.

\subsubsection{Step 3: Building an Interactive Research Prototype}

\begin{figure*}
\begin{minipage}[t][9.4cm][t]{0.7\textwidth}
  \centering
  \includegraphics[width=\textwidth]{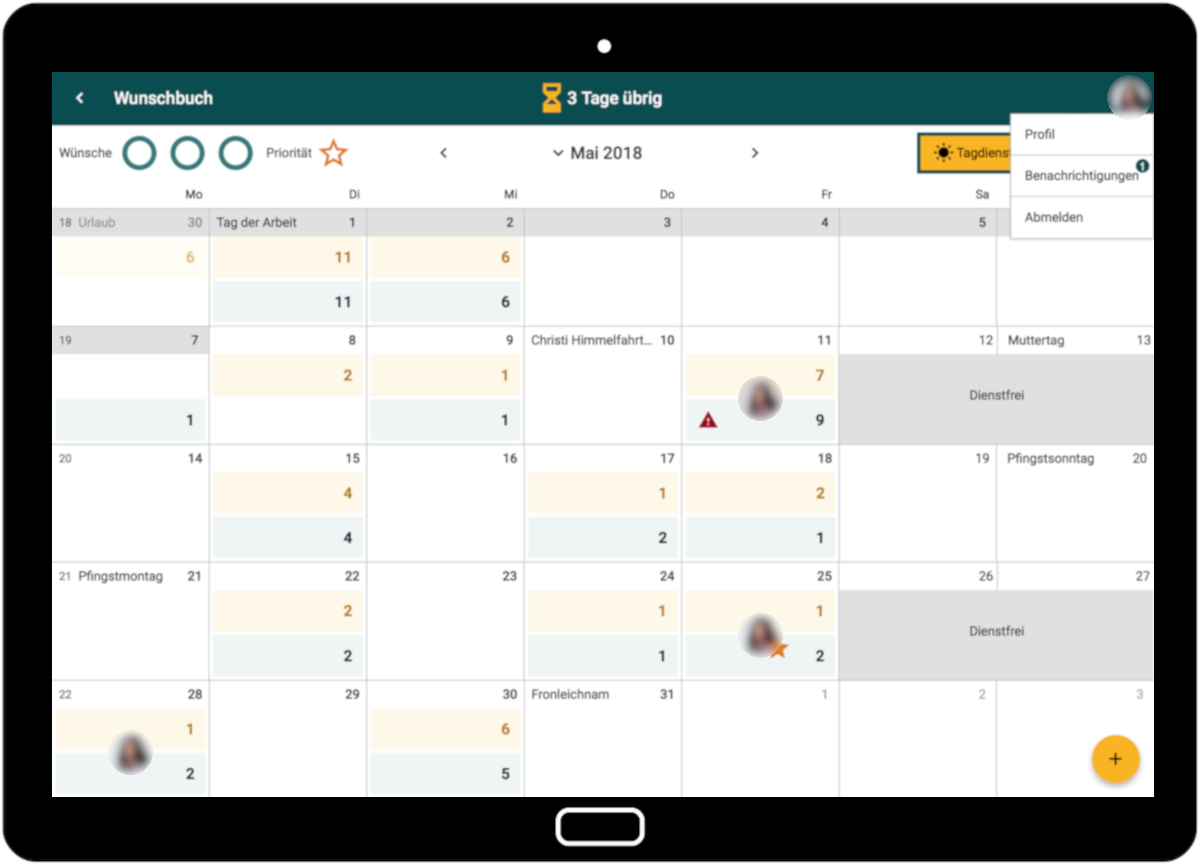}
  \subcaption{Calendar overview representing the ``safe harbor'' practice and showing how many wishes have been submitted on which day. The own wishes are marked with the profile picture and remaining conflicts are indicated with a red triangular warning sign.}
  \label{fig:calendar}
\end{minipage}

\begin{minipage}[t]{0.4\textwidth}
  \centering
  \includegraphics[width=0.9\textwidth]{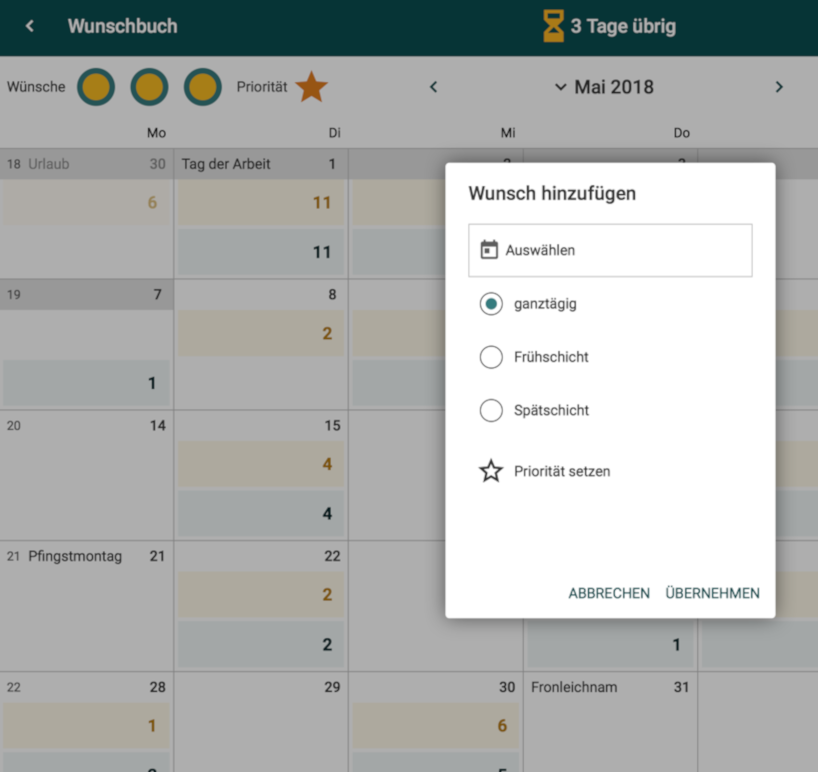}
  \subcaption{The ``wishes'' practice in the user interface. The dialog allowed to enter wishes with the options to select a date and whether it is for a morning shift, an afternoon shift, or the whole day. The ``priority'' button was removed after the first evaluation.}
  \label{fig:wishes}
\end{minipage}\hspace{0.05\textwidth}%
\begin{minipage}[t]{0.4\textwidth}
  \centering
  \includegraphics[width=0.9\textwidth]{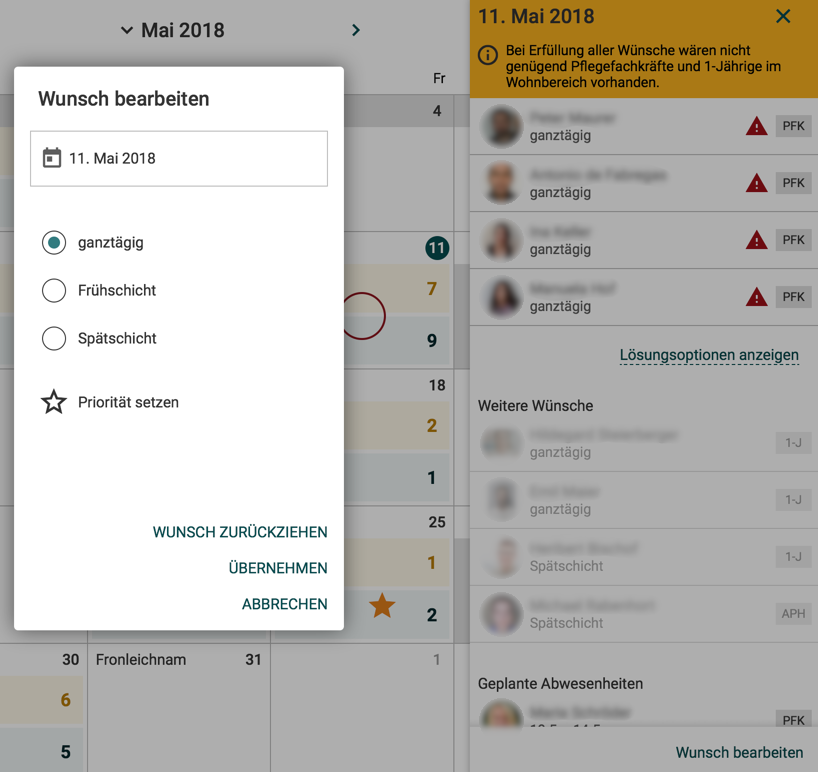}
  \subcaption{The ``conflict hero'' practice in the user interface. It contains a dialog on the left for conflict resolution, allowing the user to change or withdraw the wish. The menu on the right contains information about involved colleagues and a button to show possible solutions.}
  \label{fig:conflicts}
\end{minipage}

\Description{The figure contains 3 screenshots of the app.}
\caption{The user interface of the prototype that was used in the subcommunity over nine months.}
\label{fig:ui}
\end{figure*}

The eight self-scheduling practices served as the conceptual frame for our interactive prototype. Together with our project partners, we implemented a minimal planning process, including three of the four practices from Phase 1, with an android tablet app. The user interface is depicted in Figure~\ref{fig:ui}. While ``wishes'', i.e., particular requests for specific free shifts, were implemented, ``preferences'', i.e., a more long-term individual pattern (e.g., ``I generally prefer morning shifts'') was not, because it required a non-trivial amount of development work while we expected only minimal interaction when submitting the preferences once. Instead, we relied on the subcommunity leader to manually take care of preferences in the background. ``Conflict hero'' and ``safe harbor'' were also implemented. The app was iteratively developed in a user-centered process, with several smaller prototypes to test the usability of the specific functions. Conflicts were calculated on an external back-end server, because the calculations were computationally intense.

``Wishes'' (Figure~\ref{fig:wishes}) could be submitted for a free morning shift, afternoon shift, or the whole day, and a ``priority'' could be given to the most important wish in the month. This was meant to facilitate discussions about other, ``non-priority'' wishes while providing predictability of the most important ones. On weekends, only wishes for a morning or an afternoon shift could be submitted and only on one's work weekends. We included this restriction because free weekends were distributed in the retirement community beforehand (normally every second weekend is free). Each worker had three wishes per month, which they considered an acceptable amount and which prevented overly egoistic planning. We decided to provide no means to give a justification why one submitted a wish, at least not in the system itself. This contrasted the paper notebook used in the previous system. In doing so, we followed the recommendation by Uhde et al.~\cite{Uhde2020a} to support face-to-face interactions during conflict resolution, which allows for a need-based process and helps overcome the dilemma of privacy and procedural fairness~\cite{Colquitt2001, Colquitt2015b}. In shift scheduling, when a wish for a free shift cannot be granted, the reasons are often private (e.g., a co-worker has a doctor's appointment). Face-to-face interactions give healthcare workers full control of which details they provide to whom, while promoting fairness of the decision-making process during conflict resolution~\cite{Uhde2020a}.

``Safe harbor'' was represented through the calendar view (Figure~\ref{fig:calendar}). Ideally, the plan should be published several weeks in advance, but due to management decisions of the retirement community and economic pressure, the planners endeavored to publish the plan two weeks ahead. As described above, weekends were planned earlier. Similarly, the wishes could be entered early on and the planning activities of co-workers could be followed, including their names, because they needed these information in case of a conflict.

For the ``conflict hero'' practice (Figure~\ref{fig:conflicts}), (only) the involved healthcare workers were notified about their conflicts after logging in. They could see who was involved and withdraw their wish through the interface, or else approach the colleagues personally. If more than one co-worker was involved, all possible solutions were shown, in case, for example, more than one worker would need to withdraw the wish.

The subcommunity leader had an additional, separate interface to finalize the schedule, that is, to include preferences for morning or afternoon shifts after wishes and conflicts had been integrated.

\section{Appropriation Study}

\subsection{Study Setting and Method}

We introduced the prototype mid-February 2019 in one of the five subcommunities that was selected because of its good team climate (i.e., no major interpersonal conflicts according to management) with a workshop explaining its functionality. In the beginning, fifteen people worked in the subcommunity, one of whom left during the study period. Moreover, the subcommunity leader was reassigned to a different house of the same employer in October because of an organizational restructuring and a different subcommunity leader took over. The relationship of the workers with the new leader was also generally described as positive. Our app was installed on an android tablet which was placed in the subcommunity.

Given the early stage of development, we were confronted with several decisions to make. First, in order to assert a functional schedule and integrate it with the other schedules in the retirement community, the previous scheduling system was still running in parallel for the central planner and subcommunity leader to use. While this may have created a tension between the two planning processes, it allowed us to collect real scheduling experiences of the workers and prompted a direct comparison of the two systems, facilitating reflection. The subcommunity leader and researchers asked the healthcare workers to use the tablet app for planning and to consider the previous process to submit wishes in the notebook only as a fallback. Second, the central planner was legally responsible for scheduling in the retirement community and he was therefore allowed to make changes to the schedule, although he agreed to keep this to a legally required minimum. Third, we only installed the software on the tablet that was placed in the subcommunity, rather than having the employees use their private devices for planning. This assured that planning took place only at work, which simplified the separation between work and free time as well as the integration of employees who have no smart devices, and furthermore reduced development effort significantly. Given that we focused on schedule creation (Phase 1), this decision did not interfere with planning practices, i.e., calling people at home was not necessary.

First wishes could be submitted for March 2019. Early in April, after the first two monthly planning cycles, we organized an early evaluation session including one focus group with the subcommunity leader and the central planner (``FG1'') (both male, $35$ and $43$ years old), as well as one with the healthcare workers without the subcommunity leader (``FG2'') ($10$ participants, $1$ male and $9$ female; $21$ to $55$ years old, $Median = 44$). Both sessions lasted for around one hour. The interview guide contained questions about technical problems and experiential outcomes, for example, how they used the ``wishes'' and ``conflict hero'' function. Based on feedback from these focus groups, the amount of ``wishes'' per healthcare worker was increased to five and the ``priority'' function was removed because the workers said that usually all wishes are of high priority. After the two focus group sessions we remained in contact with the subcommunity leader(s) during the study phase, sent reminders for them to motivate the team to use the system, collected feedback on technical or usability problems and fixed them timely.

We conducted a second evaluation session in November 2019 with seven individual interviews, including one with the new subcommunity leader. The participants were between $21$ and $55$ years old ($Median = 30$) and had between $3$ and $37$ years of experience working in healthcare ($Median = 8$). All participants were female. The interview guidelines were created following the ``Interpretative Phenomenological Analysis'' (IPA) method~\cite{Smith2009}. IPA is an idiographic approach to capture subjective experiences. We chose this method because our primary interest was the appropriation of the system from the workers' point of view and we probed experiences relating to our intervention that were perceived as meaningful, those that were not, and why. The interviews lasted around 35 minutes on average. The seven individual interviews from November and the two focus groups from April served as the central data corpus for this study, which include subjective experiences of both subcommunity leaders and most of the healthcare workers. They were audio recorded after taking informed consent and transcribed in German.

Our analysis followed the IPA process~\cite{Smith2009} as well. In a first step, two independent coders (authors one and two) read the transcripts and listened to the audio files of the two focus groups as well as the seven individual interviews while taking notes on linguistic and conceptual patterns. They inductively identified themes for each interview and focus group. Second, they merged the themes across the interviews and focus groups when similar patterns crystallized in more than one case. Third, the two coders compared their two theme sets, identified common patterns between both analyses, and established consensus about the interpretation.

Complementing the subjective account of the healthcare workers, we first briefly present basic, quantitative usage data that were collected just before the second interview sessions early in November. These serve as an objective reference point for our main, qualitative analysis which is organized around the central themes of the interviews, focuses on the workers' subjective experiences with our system, and how our intended practices unfolded in the field.

\subsection{Self-scheduling in Practice}

\subsubsection{Usage Data}

\begin{figure}%
\begin{minipage}[t]{0.4\textwidth}
  \centering
  \includegraphics[width=\textwidth]{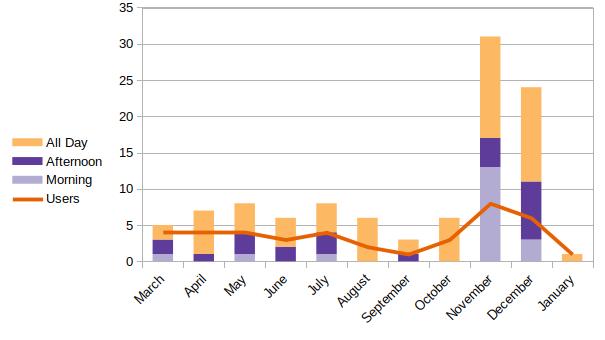}
  \subcaption{Amount of wishes per month, with the peak in November. The horizontal line represents the number of users who submitted wishes in that month.}
  \label{fig:wishes_per_month}
\end{minipage}%
\hspace{0.05\textwidth}%
\begin{minipage}[t]{0.4\textwidth}
  \centering
  \includegraphics[width=\textwidth]{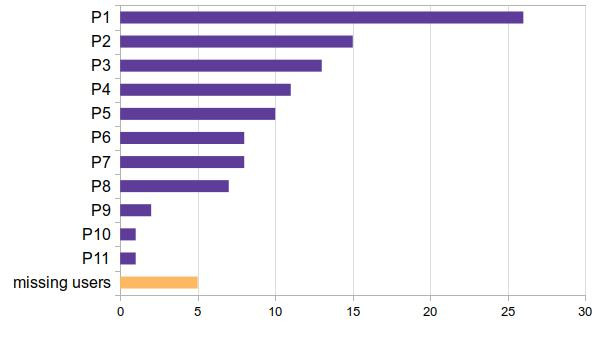}
  \subcaption{Amount of wishes per worker. Five people did not submit any wishes, while the most active user submitted 26.}
  \label{fig:wishes_per_worker}
\end{minipage}%
\Description{Two bar graphs showing the number of wishes submitted for each month from March to January (graph 1) and per worker (graph 2). Graph 1 shows that workers submitted between 3 and 8 wishes per month, but a peak of 31 and 24 wishes in November and December, respectively. Graph 2 shows that one worker (P1) submitted most wishes (26), while most other workers submitted between 7 and 15 wishes. Three workers submitted between 1 and 2 wishes, and 5 workers submitted no wishes.}
\caption{Distribution of wishes by month and worker during our study.}
\label{fig:user_data}
\end{figure}

In total, $105$ wishes (i.e., requests for free shifts) were submitted to the system by $11$ healthcare workers (see Figure~\ref{fig:user_data}). Five workers did not use the system and the subcommunity leaders only used it until/from October. Nineteen wishes were submitted for free morning shifts, $24$ for afternoon shifts, and $62$ for the whole day. During our analysis, we noticed that a lot more wishes were submitted for November and December, compared with the other months. Upon request, the new subcommunity leader explained to us that she had used a different planning procedure for November, just after she had joined the group. She had asked some workers to directly enter their wishes through the planner's separate interface, together with her. This interface was intended only for the subcommunity leader and central planner. It allowed them to adjust the preferences and individual shifts after all wishes had been entered, to account for the legal responsibility of the central planner. Therefore, one user had ten wishes in November which was usually not possible. Excluding November, $74$ wishes ($6$ morning shifts, $20$ afternoon shifts, $48$ whole days) were submitted. The peak in December is due to Christmas and New Year's Eve, which summed up to eleven wishes.

There was a distinctive disparity between different individuals and their system use. Five out of sixteen team members submitted more than one wish per month on average. The most active user alone accounted for more wishes than ten of her co-workers taken together. The reasons for this disparity are detailed in the following, qualitative analysis.

Data collection has also been activated for conflict resolution, but the feature had not been used at the time of our study, despite our explanation in the introductory workshop. The reasons for this will also be clarified below.

\subsubsection{Appropriation of Individual Planning Practices: What is `Worth a Wish'?}

We intended the ``wishes'' practice to allow for a higher quality in the workers' private lives and increased autonomy, for example, by allowing them to plan for enjoyable leisure activities. This was only partially what we found. In the subcommunity, opinions differed among the workers about what kind of wishes they would submit, resulting in the large disparity in the number of submitted wishes. There did not seem to be a general agreement on what is ``worth a wish'' and what is not. A clear case were doctor's appointments and weddings. However, less institutionalized activities varied in subjective importance and in whether or not they were submitted. When asked about the reasons for their wishes, one participant said: \textit{``For example `birthday of my boyfriend', I have submitted that in the app and in the notebook''} [I3; 32]. Another one who told us that she only sees her partner every second weekend, submitted only one wish during the entire period (to both systems). The subjective importance of the weekends spent together became clear throughout the interview, so we asked why she did not plan for them.

\begin{quote}
Participant: \textit{``That's always difficult because, like I said, the distance. And, well, in order to go to the Christmas market or so I wouldn't submit a wish.''}\\
Interviewer: \textit{``Okay. How do you think your co-workers would react if you submitted such a wish?''}\\
Participant: \textit{``I think that wouldn't be a problem.''}\\
Interviewer: \textit{``But you still wouldn't do it?''}\\
Participant: \textit{``No... I think I'm too proud for that (laughs)''} [I1; 194-198]
\end{quote}

Thus, although she had only very limited time to spend with her partner, and although she expected that her wish would not be seen as problematic, she did not consider their plans important enough to ask for free time in the group beforehand (even without having to submit a reason). This contrasts the typical understanding of workers trying to maximize their individual ``utility'' by requesting and competing for free time (e.g.,~\cite{Constantino2015, Lin2015}). Instead, this healthcare worker prioritized her reputation within the team.

A further important prerequisite for submitting wishes was that the workers were willing and able to plan for them, both in terms of their individual lifestyle and the nature of the wish. One participant said: \textit{``Actually I don't use so many} [wishes]\textit{. Because in the end I'm fine with it... I usually only use a wish if I know specifically: ‘I want to go there, you know, because I need free time, I don't know. I want that birthday, I want to go to a wedding, to a concert. Then I know it. But normally, when I have, let's say, no appointments in a month, I don't make a wish. Then I just wait and see what's coming''} [I2; 202].  In contrast, another team member said: \textit{``Usually I check it when I'm home, which appointments I have during the next two months... and then at some point I look at what's happening in January. Before December comes, then January, February, and then I submit it for the next two months''} [I7; 322]. This was positively referred to by one of her co-workers: \textit{``Well, she is someone who has regular wishes. Because... she is just a person that... she can already plan a lot for the future. I can't manage that yet. I don't even know what I should eat next week. I always have to go grocery shopping day after day, and she just has it all structured. Then she knows: `Okay, I have appointments then and then. Here is where we want to go on a trip'. And then she plans it in advance. So that's nice for her with the app, because she can sort of make an overview for the year''} [I3; 92]. There was no grudge involved about the colleague submitting more wishes, but rather a mild form of admiration.

Taken together, submitting wishes requires courage to demand free time if the wish itself provides no obvious justification (such as a wedding). Moreover, people differ in how much they plan ahead, which resulted in the different amount of wishes between different healthcare workers that we found in the usage data.

The fact that no justifications could be provided in the app sparked some discussion. Some participants assumed it was a still missing feature that was going to be added later on. Thus, some used the paper notebook additionally, in order to give a reason for their wish, but reflected on it:

\begin{quote}
Participant: \textit{``Strangely enough, I do} [submit a reason in the paper notebook]\textit{. Although I don't need to. I mean, it's of nobody else's concern why I want that day off. But somehow I still do.''}\\
Interviewer: \textit{``Why?''}\\
Participant: \textit{``So that it's clear that it's important. And that it should be taken into consideration. Because maybe I write: `I want to go to a concert', but the other one has an appointment with the doctor. So he says: ‘the doctor is more important than your free time'... and, well.''}\\
Interviewer: \textit{``But in that case it's not good to write that you want to go to a concert, is it?''}\\
Participant: \textit{``Yes, not in that case. But if it's a birthday or so, if I really want to have the time off, then I do write it.''}\\
Interviewer: \textit{``Okay, but that would mean that the concert is somehow less important...''}\\
Participant: \textit{``For others, certainly. For me? No''} [I2; 206-214]
\end{quote}

Another participant indicated a certain peer pressure: \textit{``Yes, I've seen, others write it. Well, then I also write it there. And then maybe... it} [the wish]\textit{ will be noticed, you know?''} [I6; 232]. These two statements reveal the problem of justifications from the perspective of the workers. Both try to emphasize that their appointment is important, but are aware that a simple comment in the notebook may not convey that message as clearly as it should. If they want to emphasize the importance, they feel an urge to justify it. But if they do, it may not be understood as important. We understand this as an indication that our envisioned face-to-face conflict resolution could help healthcare workers resolve that problem in a better way, although it was not adopted by all yet.

Although we designed the system with peer-to-peer conflict resolution in mind, the above statements also imply that the workers still assumed a ``central judge'' to make critical decisions. We have asked the new leader about it: \textit{``Nobody needs to tell me why they want time off. For me, that's... a wish is a wish. That has to be given. That's what I think about it. They don't want the time off without a reason. And if they don't get it, well, you know, some folks, they just don't show up. And that's what one wants to avoid''} [I4; 178-184]. And the previous one:  \textit{``If I had a colleague who runs to the doctor twenty times a month, then I would be wondering: ‘Is that right? And others have to swap their shifts for that?'... It \textbf{is} interesting... but actually it's not of my concern''} [FG1; 481-484].

While both leaders described themselves as very open about wishes, workers reflected on having to tell them about their plans in the previous system: \textit{``Well, I think... because if you always have to go to the ward leader... that is... it's very difficult. I'd have scruples, honestly... to bother him every time''} [FG2; 341-344], and: \textit{``Yes, well, we just thought... well, maybe some don't dare} [submitting wishes]\textit{. We just... had this case that you don't want to put this burden on the ward leader to make the decision and so on. And it would be nice if you could also... submit some half important appointments, if that doesn't bother anyone''} [FG2; 750-756]. These statements are in clear support of the self-scheduling process, reducing feelings of guilt that can prevent healthcare workers from submitting wishes.

Finally, although justifications were mostly used to signal that a wish is important, one participant also suggested the opposite: \textit{``In my case it's usually about the kids or appointments with the doctor or so. But well, if I actually have some... uhm... private party or so, I could write it down. So that the other one sees that it's not that important, that we can discuss about it''} [I5; 96].

Taken together, workers had the freedom to use more wishes, but some did not do so for several reasons. First, submitting many wishes may be seen as a risk to one's reputation in the group, although from the perspective of a colleague it was seen rather positively. Second, different lifestyles allow for different amounts of wishes. And third, the justifications are used as a signal of importance and thus imply that wishes should only be used in exceptional cases. Our ``wishes'' function, which provided no means to leave a justification, has been used to that end successfully and sparked reflections about the role of justifications.

\subsubsection{Conflict Prevention Behind the Scenes}

At first glance, conflicts did not seem to exist. The (lack of) usage data showed that the conflict resolution process was not used as intended, despite our introductory session explaining the feature. We inquired several times and during both feedback sessions, with similar replies: \textit{``No, I didn't... we didn't have that yet''} [I5; 102], \textit{``No, there was nothing''} [I6; 72], \textit{``And, actually, there haven't been any conflicts''} [I7; 66].

However, we noticed that several informal strategies were used to manage and avoid conflicts before they arise. For example, some participants checked if someone else had submitted a wish before planning a private appointment: \textit{``Then you can also, if you see: ‘Ah, someone else has already submitted something', and then you can pick another date or so. It gives a better overview''} [I5; 80-82]. In another case, a healthcare worker who had submitted wishes was aware of her cooperative colleagues: \textit{``Everyone knows: There are the small children. So they are so friendly to pick another date''} [I5; 108]. Thus, in a first step, the healthcare workers avoided conflicts by rescheduling their private appointments, if possible.

If that did not work, they did submit a wish on the same day as a colleague, but that was often unproblematic. In everyday life, there is enough staff in the subcommunity so that even two certified nurses can take the same day off. Sometimes, however, the situation gets more difficult, for example, during weekends or if one of the certified nurses is on vacation. Nevertheless, these cases can often be resolved without a ``real'' conflict by activating further resources, e.g., from the other subcommunities: \textit{``Yes, if they are four people} [in another subcommunity]~\textit{, then maybe someone can come down and help us a little or so''} [I4; 248], or else: \textit{``The nurses with a one year apprenticeship can also take over a shift lead, if there is a certified nurse on a different ward as a supervisor''} [I3; 207]. Moreover, they would try to minimize the impact of their wish: \textit{``If I don't get it, that would be inconvenient. In that case I could do only a} [irregularly] \textit{short morning shift. That is the alternative''} [I6; 66]. The second informal practice was to activate further resources, outside of the restricted scheduling system. However, these resources are only meant for exceptional cases and not to be used regularly.

Sometimes, despite all precautions, a scheduling conflict cannot be avoided. This is common on traditional holidays, in our case e.g., Christmas and New Year's Eve, which were planned in November. Fortunately, these are predictable and the subcommunity has established special precautions in the form of informal rules: \textit{``Yes, well... but we have the rule: Those who work on Christmas get the New Year's shifts off and vice versa''} [I4; 86]. This rule serves as a general frame of reference for everyone. The subcommunity leader had put a list in the office early on so that everyone can say which holiday shifts they prefer. On top of that, some have submitted wishes in our app. We found proactive, pro-social strategies that eased the tension: \textit{``We're lucky here to have a Muslim team member and she has directly said: ‘I'll do the late shift on Christmas, because I don't celebrate it'. And then another one has volunteered with no children, no family. So it was directly clear who was going to work''} [I4; 128-130]. This doesn't always work, however: \textit{``New Year's Eve was more difficult this time.} [...]\textit{ So I mentioned it in the team meeting, I said: ‘I need a certified nurse in the morning between New Year's Eve and on the 1st of January.} [...] \textit{And then somebody volunteered. And she said: ‘Okay, I'll do it.', but she has asked to get the 2nd of January off''} [I4; 110]. Notably, the subcommunity leader played a central role in solving this conflict by initiating the resolution process. However, she did not make the decision herself and instead worked towards a decision coming from the team.

These rules were of immense importance during conflict resolution. In fact, they surpassed a planning logic of efficiency or sufficiency. For example, one participant insisted on her late shift on 1st of January, even if she were not needed:

\begin{quote}
Participant: \textit{``But now we are three, all of whom want to work. And nobody wants to leave the afternoon shift, because we want to sleep off after the party and then go to work. And there is one person to spare.''}\\
Interviewer: \textit{``Okay, but if the early shift is already taken...''}\\
Participant: \textit{``Yes, then we have to see if one person gets the day off. But that wouldn't be right, because that one already has Christmas off. It's either the one or the other... If you're free on Christmas, you work on New Year's Eve and the other way round. And that person cannot get both days off. So he/she needs to go somewhere else''} [I2; 128-130].
\end{quote}

Finally, sometimes there are planning conflicts that are less predictable and no simple solution can be found. One such case was a staff outing to the Oktoberfest in Munich. Several subcommunity members wanted to participate. Unfortunately, the trip overlapped with the school holidays, so one team member had taken vacation already. The (previous) subcommunity leader and two more certified nurses wanted to go, but because people from the other subcommunities also participated, the situation became difficult. The wish could not be moved to a different date, external resources were depleted, and no rules had been established beforehand for that specific case. In the end, the general manager of the retirement community made a decision. She assigned the subcommunity leader for the trip, because he was in the transition to another house at that time and it was understood as an informal opportunity to get to know the new team. Then she drew lots between the other two. The one who lost said: \textit{``Well, that's how it was. I have accepted it quickly. I also thought: ‘Okay...} [...]\textit{ well, only} [the colleague]\textit{ \textbf{or} me. And then I thought, well, without her, I don't want to go anyway.} [...] \textit{Maybe it's better this way, because you don't have your friends and family with you. And it would have felt weird, if she} [...]\textit{ had stayed home''} [I3; 146]. In the end, the ``winner'' called in sick that day and did not go either. Instead, an apprentice nurse ``stood in'' for the trip.

In sum, the pro-social conflict prevention process worked mostly independent of our technical function, but followed the same fundamental principle: If conflicts cannot be avoided altogether, they are resolved face-to-face, based on individual needs first. If that provided no satisfactory solution, it was followed by an equality-based mitigation process. Only in one case where this process had also failed and which involved the entire retirement community, a top-down decision was made. The several levels of escalation illustrate that the healthcare workers try to find acceptable solutions for everyone and use all available options, even some that may be outside of the scope of a typical scheduling system. With regards to our interactive system, the information about the time and date of everyone's wishes proved useful.

\subsubsection{Early Planning of Important Appointments}

Most workers waited impatiently for the plan to be released: \textit{``Well, first everyone rushes directly to the new schedule, because you want to know how you work the next month. So that you can directly start planning''} [I3; 162]. It is important for both preparing private activities and work: \textit{``I usually look at the weekends first, if they are correct. And the second thing is if I see: ‘Okay, there is a big block!' And then I count how many days in a row I work. So that I can prepare for that''} [I3; 173].

Some reported that other employers release the schedule earlier: \textit{``When I was in the hospital with my daughter, she} [the hospital nurse]\textit{ told me: ‘What do you mean, you don't know it yet? We always get it in advance!' ''}, \textit{``That's why I say, at least more than a month so that you can plan a little bit. Because, as I said, on April 15 I theoretically don't know how I work on 1st of May... because it's not released.''}, \textit{``That makes the doctor's appointments and everything so difficult''} [FG2; 183-193]. Some healthcare workers start planning even important private activities only after the schedule is released, which should thus happen as early as possible. We are aware that the ability and willingness of the management to plan earlier depends on several external factors. In our case, the retirement community receives financial support for each resident. When a resident passes away or leaves the community for other reasons, financial support stops on that day, while the company has to pay for the scheduled staff until another resident moves in. So the management needs to consider the risk of overplanning. The short planning cycles are a direct result of this economic pressure, which is different in e.g., hospitals where the number of patients has to be estimated differently during planning. However, we think that a more fine-grained system that reduces economic risks for the management on the one hand, and allows workers to plan for punctuated, important appointments on the other, is possible.

Taken together, for subjectively important appointments, early release of the schedule facilitates planning but puts the company at a higher financial risk of overstaffing. A compromise between flexibility and the option to plan for individual appointments would help provide security for both the workers and the company. The simple calendar overview our system provided fulfilled the need to check one's appointments. But the release date turned out to be the most important factor.

\subsubsection{Leadership in Self-scheduling}

Self-scheduling as envisioned by our process assumes no designated leader and each team member has a similar power over the schedule. Nevertheless, we found several important functions of a leader, beyond being the decision-guide during conflicts, that need to be accounted for. In our case study, the workers did not plan in an exceedingly selfish way, but rather stood back with their individual wishes more than needed. Both subcommunity leaders urged their team members to submit wishes, and motivated them to think about occasions when they could take a day off for some leisure activity: \textit{``Yeah, well... because they don't have so many wishes I had to push them a little. ‘Come on, submit a wish! Look!'. And then they accepted that''} [FG1; 184-186]. The newer subcommunity leader also reported to have established a new, worker-oriented rule: \textit{``I had an afternoon shift with her on a Sunday and I already felt that something was wrong. And a few days later she told me that Sunday had been her birthday. So I said: ‘What do you mean, we \textbf{worked} together!', ‘Yes, but I didn't dare saying it', because she had already swapped that} [another]\textit{ weekend.} [...]\textit{ And I was very shocked. And since then we have a birthday calendar} [...]\textit{ I felt really sorry, also because she has two small children and, well...} [...] \textit{And it's really important for me that nobody needs to come to work on their birthday''} [I4; 274-290].

Both leaders also made several adjustments to the schedule that are not obvious for an outsider. Some were based on individual considerations: \textit{``Well, you know your team. One only wants to work up to four days in a row. Others say: ‘I'd prefer a block of seven days' ''} [I4; 210], or team composition: \textit{``This one doesn't like to work with that one, that other one not with him. But I also have to see that the combinations are good. That I have a strong one here, and a strong one there. Not that all the strong ones are in one weekend. That doesn't work. Because then the other weekend will go down''} [I4; 446].

Moreover, group-level fairness played a role in these adjustments. One team member had a part-time position with only few hours to work every week. So in the past she had not been assigned to weekends but instead stood in for sick team members more often. The new subcommunity leader changed that: \textit{``That's where I say: ‘I don't do that'. Because it's also unfair for the others. Because she had some months where she had not worked a single weekend. And there were others who said: ‘Well, why doesn't she work now, yet again?'. And I don't think that's right. In that case you need to plan everyone. And also: I prefer to plan all employees I have for the weekend. And if someone is sick, you can still use the ones that are available. I can say: ‘Hey, look. We have six men in the morning shift. We can give some away' ''} [I4; 406-412]. This adjustment to assign everyone equally was not necessarily better from an efficiency perspective. But it was important for the subjective experience of fairness.

As a team member, the subcommunity leader was accepted to have a say on the schedule. This was not always the case for the central planner, who was often complained about. Being an ``external'' person, his decisions were considered somewhat careless. For example, the subcommunity leader once had submitted a wish that had been replanned: \textit{``And then I went to see him. And asked what that was for. So he said ‘Well, it would be better like that'. And I said ‘But I submitted a wish there!' But he can't know that. So I said ‘And that's exactly the problem. Ask before you change something, otherwise it doesn't work''} [I4; 308].

In sum, the role of a subcommunity leader was still valuable despite the reduced responsibility in planning. First, they motivated the colleagues to submit wishes. Second, they facilitated conflict resolution. And third, through smaller, informal practices of team composition and by having an eye on the private lives of the workers, they adjusted the schedule, making it more agreeable and fair for the team. In a leaderless self-scheduling system, these functions need to be accounted for differently. We offer some possibilities below.

\section{Central Findings and Discussion}

In this article we presented a case study about the design and appropriation of a computer-supported collaborative self-scheduling system. Over the course of nine months, we found complex, sociotechnical interactions between informal social contracts, individual values, and socially aware planning practices that shaped how healthcare workers engaged in self-scheduling. We will first discuss our central findings about self-scheduling practices and how our app could support them, based on the themes of the previous analysis. For each theme we also add a note on how it relates to psychological need fulfillment and thus to subjective well-being. Then we discuss our findings in the broader context of related literature. And finally, we present design guidelines based on our insights for computer-supported self-scheduling.

\subsection{Central Findings}

\subsubsection{Individual Planning Practices in Context} 

Concerning individual planning, we found evidence that is inconsistent with a simplistic view on individual scheduling as a means for the healthcare workers to maximize individual ``utility''. When using the ``wishes'' practice to schedule individual free time, workers were aware of their team as a social group and their standing within it, and they used their freedom to plan in a responsible way. Differences in lifestyle, i.e., whether one plans private appointments in advance or lives more spontaneously, led to a disparity in the usage of wishes but not to grudge. A central insight is that, rather than problems relating to overuse of this individual freedom, some workers may be too careful to claim specific free shifts on a regular basis. Moreover, we illustrated the dilemma of justifying wishes in written form, which failed to convey their subjective importance. This supports our notion to foster face-to-face interaction. Overall, individual self-scheduling practices were successfully supported by our scheduling app.

Regarding need fulfillment, the ``wishes'' practice was primarily designed to address autonomy which has worked relatively well by allowing workers to engage in planning. Two hurdles we found related to reputation and different lifestyles. We think that reputation can be a threat to autonomy for some healthcare workers and should be addressed in the design. The differences in lifestyle are not, and in fact they can rather be understood as the expression of autonomous decision-making in private, which results in these different lifestyles. In our app, the general functionality to plan more autonomously was provided. Design guidelines that could help overcome the reputation problems are presented below.

\subsubsection{Pro-social Conflict Prevention} 

Conflict prevention and resolution were similarly embedded in the social context. Our participants had several pro-social informal practices, resulting in a step-wise, escalating conflict resolution process. First, workers tried to avoid a conflict altogether by rearranging their private plans. Second, they searched for external resources to find an acceptable solution for all involved parties, based on personal needs. Third, for predictable conflicts, they relied on established, equality-based rules to find a solution. In one case, all precautions failed and the conflict was escalated to the general management that made a workable but unsatisfying decision. Being based on a too abstract understanding of conflict resolution in self-scheduling, our app did not appropriately support the conflict prevention process.

In terms of subjective well-being, the ``conflict hero'' practice was designed to support the need for popularity (e.g.,~\cite{Hassenzahl2010a}). The central idea was that participants respect each other's needs and work together to find a common solution, thus fostering a positive team spirit. This corresponds to our findings of a pro-social conflict prevention process.

\subsubsection{Private Planning on Hold} 

Early release of the schedule proved to be essential for some workers to plan private appointments. Some of our participants tended to wait until the schedule was released before they made private plans, e.g., with the doctor. Thus, private life was on hold until the work schedule was set. Our app was limited in its support for early planning, given the economic pressure on the management particularly in German retirement communities. An important takeaway of our study is that the release date can be seen as a means of ``negotiation of flexibility'', with respect to other societal actors such as clinics, kindergartens, the friends and partners with their work times, or cultural events. A late release means that the workers have to reschedule and possibly solve private planning conflicts with their co-workers on short notice, which could be avoided if they knew the schedule earlier. Taking this into account, we understand work and private life not as fundamental opposites, but rather as co-emerging processes. While it may be counterintuitive to advocate a ``less flexible'' (i.e., decided early on) work schedule in order to support private life, the high pressure of the work schedule on private life is independent of its release date. Early (partial) release facilitates informed private planning.

Accordingly, the associated ``safe harbor'' practice could only fulfill the need for security to a limited extent. Once the schedule was released and workers saw their individual schedule, they did not only verify whether their private appointments could be met, but also anticipated phases of high workload to plan for them ahead of time. Taken together, the need for security was central to self-scheduling, but directly affected by external decisions.

\subsubsection{Leadership}

The ambiguous role of the subcommunity leader in our case study highlighted several implicit leadership functions they have in central scheduling which need to be accounted for in self-scheduling systems, which per definition have no leader. The leaders motivated their colleagues to occasionally prioritize their private lives above work. They also initiated conflict resolution. Moreover, they had an eye on team composition, group-level fairness, and implemented small pro-social rules to make the schedule more agreeable. However, there are also several drawbacks of a leader role in shift scheduling. We found that the mere practice to submit a request for free time to a leader, thereby creating additional workload for them, was seen as a hurdle. Moreover, as was found previously~\cite{Roennberg2010}, a leader could be seen as a gatekeeper and reduce subjective control of one's schedule. In terms of subjective well-being, we can only tentatively interpret the role of leadership as a moderating factor that shapes whether and how healthcare workers engage in planning, when and how conflicts are resolved, and that makes adjustments on a group level.

\subsection{Broader Impact}

This detailed account of the design and appropriation of computer-supported self-scheduling in healthcare complements previous, more abstract work~\cite{Miller1984, Roennberg2010, Uhde2020a}. While Rönnberg et al.~\cite{Roennberg2010} presented an early, inclusive scheduling system for healthcare workers, they did not detail how exactly individual and group planning practices may unfold in context. Uhde et al.~\cite{Uhde2020a} sketched an abstract, inclusive conflict resolution process based on negotiating individual needs first. It includes equality measures as a candidate for second order conflict resolutions if the needs fail to indicate a solution. In the present study we presented a detailed picture of how this process can look like in practice in a retirement community and we could confirm the overall process: Needs are used for initial conflict resolution and equality-based rules are used as a fallback.

The present study contrasts previous automation-oriented scheduling solutions~\cite{Constantino2013, Constantino2015, Lin2015} that are based on the assumption that workers approach scheduling with the goal to maximize individual ``utility'' (i.e., control of free time), which we did not find. Instead, we found high social awareness and responsible use of the individual freedom to plan. This insight can be a first step to refocus research about scheduling systems away from purely automated, conflict- or problem-oriented approaches. In fact, the only conflict that we found to qualify for automated decision-making, given that no inclusive decision could be found, was the staff outing to the Oktoberfest. During the nine month period, only one such conflict appeared. Considering that this was a very specific case that did not allow for a comparison with previous ones and that it included further actors beyond the subcommunity, we question the need for and usefulness of automated decision-making in shift scheduling. Instead, particularized, case specific justice~\cite{Binns2020} may be important here.

The long-term consequences of shift work impact healthcare workers' subjective well-being~\cite{Fenwick2001, Hulsegge2020, Kubo2013, Rotenstein2018}. Our design case exemplified how this may be countered to an extent by setting subjective well-being as a high level priority during the design of scheduling systems. This focus on subjective, positive experiences of involved healthcare workers extends previous work on hospital organization primarily focused on organizational tasks~\cite{Bardram2010, Bossen2016, Krabbel1997, Stisen2016}, patients~\cite{Kuchera2011, Moller2013, Wolf1997}, or illnesses~\cite{Strauss1985}. More broadly, our study presents a case of how digital technology can support self-organized work (e.g.,~\cite{Laloux2014, Lee2017c}). Beyond shift work, it outlines an example for how problems associated with power and inequality (e.g.,~\cite{Clear2017, DaCunha2008, Salehi2015}) can be resolved through more inclusive decision-making mediated by technology.

In an earlier study, Clear et al.~\cite{Clear2017} had highlighted the importance of different subjective perspectives among workers and management, and the role that seemingly ``objective'' but incomplete information can play in shaping work conflicts. Similarly, the justifications for free shifts in our study were submitted as ``objective'' facts. Workers anticipated their importance for later conflict resolution and hoped they would support their position on their behalf. Nunes et al.~\cite{Nunes2018} have described how shift workers make decisions that are negative for their health, e.g., when standing in for colleagues although they need a break themselves. In return, they can rely on the colleagues to help them out when they need it, which gives them security for private planning. This collective process of trust-building may have kept some nurses from submitting too many wishes, but it describes a thin line between exploiting oneself now and hoping to be supported in the future. Both studies illustrate a high social awareness during work-related decision-making and how workers try to use the means they have to influence decisions in a favorable way. We think that providing workers with control about the actual process is a promising path to make such circumventions unnecessary.

\subsection{Design Guidelines}

For future implementations of self-scheduling systems, we conclude with some central design guidelines based on our findings.

\subsubsection{Scaffolding of Individual Planning Practices}

Practices of individual planning need to be understood within the social context~\cite{Ackerman2000, Selbst2019}. While the focus of dealing with worker influence on the schedule typically lies on resolving conflicts~\cite{Constantino2015, Uhde2020a}, designers should consider social hurdles that complicate individual planning for healthcare workers, such as an expected negative impact of submitting ``unimportant'' wishes on one's reputation. A technical successor of our ``wishes'' practice could include mechanisms to counter this negative impact. One possibility would be to provide examples of acceptable wishes in the interface, such as visiting a Christmas market or a concert. Based on this, workers' might change their beliefs about what is ``worth a wish''. Moreover, these examples could provide a vague but objective reference point to support their position in possible negotiations~\cite{Clear2017}. Other possibilities include the support to share stories about ``unimportant'' wishes within the group. Based on our findings, we recommend not to include a possibility to provide detailed justifications, however, because they can misrepresent the subjective importance of a request. Abstract, vague markers (e.g., ``time for myself'' or ``I need a change of scene'') could be used alternatively and optionally for cases that are less important. In order to further support workers with spontaneous lifestyles, reminders to submit wishes or suggestions for possible wishes could be included.

\subsubsection{Dealing With Conflicts}

Conflict resolution already worked fine without technical support in most cases and, as in our system, the necessary information about others' wishes should be provided early on. But the outlined process can be augmented in several ways. First, additional resources in the specific context could be presented more readily. For example, tacit indicators, such as color codes in the calendar that indicate days with many wishes in other subcommunities, could further help anticipate and facilitate conflicts and possibly even promote pro-active collaboration across subcommunities. Second, the equality-based rules for predictable conflicts could be better integrated. For instance, in our case, the rule for Christmas and New Year's Eve could be picked up, become represented through the system, and actively addressed as a dedicated event, reducing the need of a leader to initiate conflict resolution. Third, technical support may have helped avoid the collision between the staff outing and school holidays by helping the planner of the outing to anticipate the staff shortage. However, the case in our study involved actors beyond the subcommunity, which have to be considered. Moreover, given that only a single case was found that was not satisfactorily resolved, better social mediation processes may be needed.

\subsubsection{Co-emergence of Private and Work Lives}

Providing a way for the healthcare workers to plan for important private appointments early on could help rebalance the economic risk of the organization while allowing workers to better plan for their private lives. The fixed weekends have helped in our study, but they could not cover appointments during the week (e.g., with the doctor). One compromise would be to decide for one or two working days per week and worker where they know their schedule in advance. This could provide the necessary stability to organize important appointments and free the wishes for less important ones. Alternatively, workers could be supported to conclude informal contracts within the team early on for specific shifts, e.g., asserting that they would be available during the colleague's appointment. This could be designed as a pro-social practice and provide security even if the schedule is not fixed yet.

\subsubsection{Self-Leadership}

Our study revealed positive impacts that a dedicated leader can have. In self-scheduling, this needs to be replaced by other means. One function of the leaders can be summarized as ``saving workers from themselves'' by regulating their workload and motivating them to take time off. First technological solutions have been presented already that could support self-care practices of shift workers, e.g., to regulate their caffeine intake and sleep patterns~\cite{Nunes2018}. In addition, Hassenzahl et al.~\cite{Hassenzahl2016} proposed a ``benevolent'' calendar to support well-being by actively balancing work and private life. Janböcke et al.~\cite{Janbocke2020} presented a calendar to help people with scheduling daily tasks more in line with their ``inner clock'' (i.e., chronobiology). In all three cases, available knowledge about healthy usage of time became implemented in a technology to productively challenge their users to think more about their own well-being. Besides practices of self-care in shift scheduling, practices of other-care that are not necessarily related to scheduling, but to the work context itself, could also help to cope with shift work~\cite{Uhde2020b}. A second function of the leaders was ``informed group composition''. The subcommunity leader argued that healthcare workers prefer to work with colleagues they like, rather than distributing ``strong'' workers across shifts. In a previous study, Langfred et al.~\cite{Langfred2007} found some evidence that group performance in self-organized work can suffer from personal conflicts, so sympathy should not generally be condemned as a bad heuristic. Physically exhausting tasks may be supported by robotic assistance~\cite{Riek2017}, which could further reduce this problem. Nevertheless, including and working with information about which workers are strong and reliable and which are not can lead to further ethical and privacy issues. Other, less problematic information (e.g., workers who have been in the subcommunity for a long time vs. those who have just joined), can be used by a scheduling system to improve group composition and highlight shifts where problems may arise. Where a dedicated leader cannot be replaced easily, other group-based mechanisms could be considered, including rotating leadership, consensus-based mechanisms, or no-veto-agreements (see~\cite{Laloux2014} for some practical examples).

\subsection{Limitations and Further Research}

Although our case study was set in a real-world healthcare environment, our findings are taken from a subcommunity where the team spirit was generally positive. This allowed us to provide a positive reference scenario, but, of course, insights into how such a system would be appropriated in a less harmonic setting are missing. Moreover, given the relatively small sample, we did not collect quantitative data about changes in well-being over time, which would add further insights about the effectiveness of our worker-oriented approach. Other long-term effects, including the integration of new healthcare workers in the team or the integration of self-scheduling healthcare institutions in the broader healthcare system, are also left for future research. In addition, we have not focused on the efficiency of our system, in terms of time and effort needed to create a schedule. A comparative study would help to estimate a possible management overhead of self-scheduling, or even indicate that they can be more efficient because conflicts are avoided or handled quickly and more directly~\cite{Laloux2014}. Finally, using the self-scheduling system in a setting where no fallback system is available, as well as a more complete implementation that includes all practices from all three phases could provide insights about a more complete appropriation.

\section{Conclusion}

Our nine month case study of a self-scheduling app in a retirement community drew a detailed picture of real-life shift planning practices and challenges. Nurses submitted their wishes autonomously, although they often used this function for ``non-leisure'' appointments and individual usage varied. In terms of conflict resolution, they engaged in several informal practices: First, they tried to avoid the conflict by rescheduling private appointments. Second, they tried to find an alternative solution, respecting the personal needs of everyone involved. In case of unavoidable conflicts, such as public holidays, informal rules helped to find a solution and team diversity facilitated the process. Where no rules existed (i.e., for unpredictable conflicts),  an ad hoc resolution process was used and an unsatisfying solution was found. Researchers and designers should focus on flexible ways of maximizing workers' ability to plan for private appointments to help reduce conflicts and provide security. Informal conflict resolution practices and additional resources need to be integrated in technical assistance. Finally, leadership helped to deal with group-level processes and we presented alternatives that could be used in leaderless self-scheduling.

\begin{acks}
The project is financed with funding provided by the \grantsponsor{GS501100002347}{German Federal Ministry of Education and Research}{https://doi.org/10.13039/501100002347} and the \grantsponsor{GS501100004895}{European Social Fund}{https://doi.org/10.13039/501100004895} under the ``Future of work'' programme (grant number~\grantnum{02L15A216}{02L15A216}).

We would like to thank all participants in our study for their support and Mena Mesenhöller and Kieu Tran for their helpful feedback on earlier versions of this paper.

\end{acks}

\bibliographystyle{ACM-Reference-Format}
\bibliography{bibliography}

\received{June 2020}
\received[revised]{October 2020}
\received[accepted]{December 2020}

\end{document}